\documentclass[amsmath,amssymb,nofootinbib,wasysym]{revtex4}

\usepackage{w-greek}
\usepackage{latexsym}
\usepackage{amssymb}
\usepackage{amsfonts,color}
\usepackage{amsmath}
\usepackage[dvips]{graphicx}
\usepackage{bm}

\def\virg#1{``#1''}
\def\rp#1#2{{#1\over#2}}
\def\ton#1{\left(#1\right)}
\def\qua#1{\left[#1\right]}
\def\lb#1{\label{#1}}

\def\hb#1{\hat{\mathbf{#1}}}
\def\vb#1{{\mathbf{#1}}}

\def\ber{\begin{eqnarray}}
\def\eer{\end{eqnarray}}
\def\beq{\begin{equation}}
\def\eeq{\end{equation}}

\begin{document}

\title{{Probing a $r^{-n}$ modification of the Newtonian potential with Exoplanets} }

\author{Matteo Luca Ruggiero}
\email{matteo.ruggiero@polito.it}
\affiliation{Politecnico di Torino, Corso Duca degli Abruzzi 24, Torino, Italy and INFN, Sezione di Pisa, Italy}

\author{Lorenzo Iorio}
\email{lorenzo.iorio@libero.it}
\affiliation{Ministero dell'Istruzione, dell'Universit\`{a} e della Ricerca (M.I.U.R.)  \\
Permanent address for correspondence: Viale Unit\`{a} di Italia 68, 70125, Bari (BA)}

\date{\today}

\begin{abstract}
The growing availability of increasingly accurate data on transiting exoplanets suggests the possibility of using these systems as possible testbeds for modified   models of gravity. In particular, we suggest that the post-Keplerian (pK) dynamical effects from  the perturbations of the Newtonian potential falling off as the square or the cube of the distance from the mass of the host star break the degeneracy of the anomalistic, draconitic and sidereal periods. The latter are characteristic temporal intervals in the motion of a binary system, and all coincide in the purely Keplerian case. We work out their analytical expressions in presence of the aforementioned perturbations to yield preliminary insights on the potential of the method proposed for constraining the modified models of gravity considered. {A comparison with other results existing in the literature is made.}
\end{abstract}

\maketitle

\section{Introduction} \label{sec:intro}

Almost 25 years after the first discovery of a planet orbiting a  main sequence star \cite{Mayor:1995eh}, there are nowadays more than 4000 known planets in more than 3000 known planetary systems\footnote{See, e.g., \url{http://www.exoplanet.eu/catalog/} on the Internet.}. Different detection techniques such as radial velocity, transit photometry and timing, pulsar timing, microlensing, astrometry \cite{perryman,book018}, sensitive to specific physical properties of the planetary systems, have been developed and used so far.   In general,  it is fruitful to combine different and complementary approaches to obtain a  knowledge that is as extensive and complete as possible of the key physical and orbital features of the planetary systems of interest. The large number of confirmed and potential exoplanets, besides giving us the possibility of improving our understanding of planets formation, provides a potentially viable opportunity of testing fundamental laws of physics, such as gravity models, outside our Solar System.

 Actually, General Relativity (GR) is the best available model of the gravitational interaction: its predictions  were verified with great accuracy during last century  even though challenges to the Einsteinian paradigm come from  cosmological observations \cite{2014arXiv1409.7871W,2016Univ....2...23D}. Moreover, besides the issues deriving from the observation of the Universe at very large scales, it is a well known fact that GR  is not renormalizable and it cannot be reconciled with a quantum framework \citep{Stelle:1976gc,2016Univ....2...24L}: this ultimately implies that our current model of gravitational interactions stands apart from the Standard Model of particle physics. As a consequence, there are strong motivations suggesting that GR is just a suitable limit of a more general theory of gravity we still do not know. A description of various modified gravity models can be found, e.g., in the review papers \cite{Tsujikawa:2010zza,2012PhR...513....1C,Berti:2015itd}.

However, due to the reliability of Einstein's theory, it is manifest that any model of modified gravity should be in agreement with the known tests of GR: every extended theory of gravity is expected to reproduce GR in a suitable weak-field limit. As a consequence, modified gravity models must have correct Newtonian and post-Newtonian limits and, up to intermediate scales, the deviations from the GR predictions can be considered as perturbations described by specific model-dependent parameters, whose values can be constrained by observations.
Here, we suggest that transiting exoplanets, thanks to the continuous improvement in the high precision measurements performed, could provide the opportunity of testing modified models of gravity outside the Solar System; for other works dealing with alternative theories of gravity and exoplanets, see, e.g., \cite{2010OAJ.....3..167I,2017PhLB..769..485V}. This is quite similar to what happened to pulsars systems, which have become  celestial laboratories for theories of gravity \cite{Stairs2003,kramer2006tests}.  {Actually, a recent work by {\citet{Blanchet:2019zxv}} suggests that  the general relativistic orbital precession could be detected by observing the exoplanet HD 80606b.}

In transiting exoplanets, the time span $T_\mathrm{exp}$ characterizing the orbital revolution, generically dubbed as `orbital period', which is actually measured is the time interval $T_\mathrm{tra}$ between two consecutive passages at the positions in the orbit, called transit centres, which minimize the sky-projected distance $r_\mathrm{sky}$ of the planet from the star \cite{Winn010}. From a theoretical point of view, it is unclear how $T_\mathrm{exp}$ can be mathematically modelled when post-Keplerian (pK) dynamical effects (star's oblateness $J_2^\star$, possible $\mathrm{N}$-body disturbances due to other planets in the system, general relativity, modified models of gravity, etc.) are at work in the system at hand. Possible choices are as follows \cite{2016MNRAS.460.2445I}.
The draconitic period $T_\mathrm{dra}$ is defined for a perturbed path as the time span between two successive instants when the real position of the test particle coincides with the ascending node position on the corresponding osculating orbit.
The anomalistic period $T_\mathrm{ano}$ can be defined as the time interval between two successive instants when the real position of the test particle
coincides with the pericentre position on the corresponding orbit.
The sidereal period $T_\mathrm{sid}$ is the time interval between two successive instants when the real position of the test particle lies on
a given reference direction in the sky. The latter may be, e.g., the one from which the longitudes are reckoned, i.e. the $x$ axis in the coordinate systems which are usually tied to the plane of the sky.
In a purely Keplerian case, $T_\mathrm{dra},\,T_\mathrm{ano},\,T_\mathrm{sid}$ are degenerate, being identical to the unperturbed Keplerian period. If some pK disturbing accelerations are present, the degeneration is generally broken \cite{2016MNRAS.460.2445I}. Here, using the approach described in \cite{SSS}, we consider pK disturbing accelerations starting from a general stationary and spherically symmetric spacetime which can be thought of a solution of the field equation of modified gravity model, describing the gravitational field around a point-like mass, which will be the model of an exoplanetary system. In particular, we will focus on the perturbations of the Newtonian potential falling off as the square or the cube of the distance from the central mass, and we will work out the expressions of $T_\mathrm{dra},\,T_\mathrm{ano},\,T_\mathrm{sid}$.

This paper is organised as follows: in Section \ref{sec:geo} we obtain the perturbing acceleration, while in Section \ref{sec:periods} we calculate the perturbations of the  anomalistic, draconitic and sidereal periods. Conclusion are drawn in Section \ref{sec:conc}.
%
\section{The perturbing acceleration} \label{sec:geo}
%
We assume that the gravitational field of a point mass in a generic modified model of gravity is described by the following static and spherically symmetric metric\footnote{If not otherwise stated, as in (\ref{ano2}) to (\ref{sid3}) below, we use units such that the speed of light in vacuum $c$ and the Newtonian gravitational constant $G$ are set equal to 1;
 bold face letters like ${\mathbf{x}}$ refer to spatial vectors while Latin indices refer to spatial components.}
\beq
 (ds)^{2}=\left[1+\phi(r)\right](dt)^{2}- \left[1+\psi(r)\right]\left[(dx)^{2}+(dy)^{2}+(dz)^{2}\right], \label{eq:metrica01}
\eeq
where $r=\sqrt{x^{2}+y^{2}+z^{2}}$, and $\phi(r),\ \psi(r)$ are functions depending on the mass $M$ of the source, which is the host star in our case, and, possibly, on other parameters of the theory.

We suppose that the effects of the the modified model  can be considered as perturbations of the known GR solution, i.e. the Schwarzschild spacetime. This amounts to saying that $\phi(r)$ and $\psi(r)$ must approach their GR values: in other words we suppose that in a suitable limit, the metric takes the form
\begin{eqnarray}
\phi(r) & = & \phi_\mathrm{GR}(r)+\phi_\mathrm{mod}(r), \label{eq:defar1} \\
\psi(r) & = & \psi_\mathrm{GR}(r)+\psi_\mathrm{mod}(r), \label{eq:defbr1}
\end{eqnarray}
where the GR values are given by $\phi_\mathrm{GR}(r)=-2M/r+2M^{2}/r^{2}$, $\psi_\mathrm{GR}(r)=2M/r$  and the perturbations $\phi_\mathrm{mod}(r), \psi_\mathrm{mod}(r)$ due to the modified gravity model are such that $\phi_\mathrm{mod}(r) \ll \phi_\mathrm{GR}(r)$, $\psi_\mathrm{mod}(r) \ll \psi_\mathrm{GR}(r)$.
In order to obtain the perturbing pK acceleration, we proceed as follows. First, we assume that in the given theory, the matter is minimally and universally coupled, so that test particles follow geodesics of the metric (\ref{eq:metrica01}).  Then, we consider the (post-Newtonian) equation of motion of a test particle (see \cite{1991ercm.book.....B})
\beq
\ddot x^{i}=-\frac 1 2 h_{00,i}- \frac 1 2 h_{ik}h_{00,k}+h_{00,k}\dot x^{k} \dot x^{i}+ \left(h_{ik,m}-\frac 1 2 h_{km,i} \right)\dot x^{k} \dot x ^{m}, \label{eq:geod1}
\eeq
where ``dot'' stands for derivative with respect to the coordinate time. Since  in our notation  it is $h_{00}=\phi(r)$ and $h_{ij}=\psi(r)\delta_{ij}$, we can write the perturbing acceleration $\vb{A}$   in the form
\beq
\vb{A}=-\frac 1 2 \left \{\Phi(r) \left[1+\psi_\mathrm{mod}(r) \right]+\Psi(r)v^{2}  \right \} \hb x+ \left[\Phi(r)+\Psi(r) \right]\left(\hb x \cdot \vb v \right) \vb v, \label{eq:acc1}
\eeq
where we set $\vb x= \left\{x,y,z \right\}$, $\vb v=\left\{\dot x, \dot y, \dot z \right\}$, $\hb x= \vb x/|\vb x|$, and\footnote{{Notice that in Eq. (\ref{eq:acc1}) we neglected cross-terms deriving from the product of  GR and modified gravity potentials.}}
\beq
\Phi(r) \doteq \frac{d \phi_\mathrm{mod}(r)}{dr}, \quad \quad \Psi(r) \doteq \frac{d \psi_\mathrm{mod}(r)}{dr}. \label{eq:defAB1}
\eeq
Since we are interested in the lowest order effects on planetary motion of $\phi_\mathrm{mod}(r)$ and $\psi_\mathrm{mod}(r)$, we may also neglect the non linear terms (i.e. non linear perturbations with respect to flat spacetime). To this end, we start by noticing that  to Newtonian order, it is $v^{2} \simeq \phi_\mathrm{GR}(r) \simeq M/r$. As a consequence,  in (\ref{eq:acc1}) we may neglect  the terms proportional to $v^{2}$ and to $\left(\hb x \cdot \vb v \right) \vb v$ (which is also proportional to the orbital eccentricity) and also the term $\Phi(r) \psi_\mathrm{mod}(r)$. In summary, in the weak-field and slow-motion limit the perturbing pK acceleration that we are going to consider is purely radial and is  given by
\beq
\vb{A}=-\frac{1}{2} \Phi(r) \hb x  \doteq A_{r} \hb x. \label{eq:acc3}
\eeq

We consider two kinds of quite generic perturbations, in the form of power laws. Given a constant $\alpha^{\ton{N}}$, which is a parameter deriving from the modified gravity model, the perturbations that we focus on are in the form:
\beq
\phi_\mathrm{mod}(r)=\frac{\alpha^{\ton{N}}}{r^{N}}, \quad N >1, \label{eq:pertneg}
\eeq
According to our definition (\ref{eq:acc3}), we have to following pK perturbing acceleration:
\beq
A_{r}=\frac{1}{2} \alpha^{\ton{N}} N \frac{1}{r^{N+1}}, \label{eq:Wpos}
\eeq
We notice that, for a logarithmic perturbation in the form $\phi_\mathrm{mod}(r)=\beta \log (r/r_{0})$, it is $A_{r}=-\frac 1 2 \frac{\beta}{r}$, which can be dealt with as in the case (\ref{eq:Wpos}) above.
{In particular, we will consider below the cases $N=2,3$.}

\section{The anomalistic, draconitic and sidereal periods} \label{sec:periods}
As we said before, in the purely Keplerian case the anomalistic, draconitic and sidereal periods $T_\mathrm{dra},\,T_\mathrm{ano},\,T_\mathrm{sid}$  are degenerate, since they coincide with unperturbed Keplerian period $T_\mathrm{K}=2\,\uppi\sqrt{a^3/\mu}$, where $a$ is the semimajor axis and $\mu\doteq GM$ is the gravitational parameter of the host star of mass $M$. As shown in \cite{2016MNRAS.460.2445I}, when some pK perturbing accelerations are present, the degeneration is generally broken. Since, at present, it is unclear which out of the aforementioned three temporal intervals is actually measured in transiting exoplanets, we will analytically calculate the pK corrections to all of them induced by the radial accelerations $A_r$ of (\ref{eq:Wpos}) for $N=2,\,3$, assumed as small perturbations of the Newtonian monopole. In the following, we will adopt the approach used in \cite{2016MNRAS.460.2445I} to deal with some standard classical and general relativistic pK accelerations. According to \cite{2016MNRAS.460.2445I}, it seems that the measured period $T_\mathrm{exp}$ should be $T_\mathrm{dra}$ for circular orbits, being possibly $T_\mathrm{ano}$ for nonzero values of the eccentricity $e$.

It turns out that, while it is possible to analytically work out $T_\mathrm{ano}$ without recurring to any simplifying assumptions in $e$,  only approximate expressions to the lowest orders in $e$ can be obtained for $T_\mathrm{dra},\,T_\mathrm{sid}$. For the anomalistic period, the pK corrections $\Delta T_\mathrm{ano}$ to  $T_\mathrm{K}$ due to (\ref{eq:Wpos}) for $N=2,\,3$  are
\begin{align}
\Delta T_\mathrm{ano} \lb{ano2} & = \rp{3\,\uppi\,\alpha^{\ton{2}}\,\sqrt{a}\,\ton{1+e\cos f_0}^2}{\mu^{3/2}\,\ton{1-e^2}^2},\, N=2, \\ \nonumber \\
\Delta T_\mathrm{ano} \lb{ano3} & = \rp{3\,\uppi\,\alpha^{\ton{3}}\,\ton{1+e\cos f_0}^3}{\sqrt{a\,\mu^3}\,\ton{1-e^2}^3},\, N=3,
\end{align}
while for the draconitic and sideral periods, the corrections $\Delta T_\mathrm{dra},\,\Delta T_\mathrm{sid}$ to the Keplerian period are
\begin{align}
\Delta T_\mathrm{dra} \lb{dra2} & \simeq \rp{4\,\uppi\,\alpha^{\ton{2}}\,\sqrt{a}\,\ton{1 + 3\,e\,\cos\omega\,\cos u_0}}{\mu^{3/2}}+\mathcal{O}\ton{e^2},\, N=2, \\ \nonumber \\
\Delta T_\mathrm{dra} \lb{dra3} & \simeq \rp{6\,\uppi\,\alpha^{\ton{3}}\,\ton{1 + 3\,e\,\cos\omega\,\cos u_0}}{\sqrt{a\,\mu^3}}+\mathcal{O}\ton{e^2},\, N=3,
\end{align}
and
\begin{align}
\Delta T_\mathrm{sid} \lb{sid2} & \simeq \rp{4\,\uppi\,\alpha^{\ton{2}}\,\sqrt{a}\,\ton{1 + 3\,e\,\cos\varpi\,\cos l_0}}{\mu^{3/2}}+\mathcal{O}\ton{e^2},\, N=2, \\ \nonumber \\
\Delta T_\mathrm{sid} \lb{sid3} & \simeq \rp{6\,\uppi\,\alpha^{\ton{3}}\,\ton{1 + 3\,e\,\cos\varpi\,\cos l_0}}{\sqrt{a\,\mu^3}}+\mathcal{O}\ton{e^2},\, N=3,
\end{align}
respectively. In (\ref{ano2}) to (\ref{sid3}), $f_0$ is the initial value of the true anomaly $f$, $\omega$ is the argument of periastron, $u_0$ is the initial value of the argument of latitude $u\doteq f+\omega$, $\varpi\doteq \Omega+\omega$ is the longitude of periastron, $\Omega$ is the longitude of the ascending node, and $l_0$ is the initial value of the true longitude $l\doteq f + \varpi$. Note also that $\qua{\alpha^{\ton{2}}} = \mathrm{L}^4\,\mathrm{T}^{-2}$, while $\qua{\alpha^{\ton{3}}} = \mathrm{L}^5\,\mathrm{T}^{-2}$.

From Eqs. (\ref{ano2})-(\ref{sid3}) we see that the three periods are degenerate also for  circular orbits $e=0$. In this case, if we set
\beq
T_\mathrm{ano}=T_\mathrm{K}+ \Delta T_\mathrm{ano}, \label{eq:totano}
\eeq

\beq
T_\mathrm{dra}=T_\mathrm{K}+ \Delta T_\mathrm{dra}, \label{eq:totdra}
\eeq

\beq
T_\mathrm{sid}=T_\mathrm{K}+ \Delta T_\mathrm{sid}, \label{eq:totsid}
\eeq
we may estimate the relative impact of the perturbation of the Keplerian period:

\beq
\frac{T_\mathrm{ano}-T_\mathrm{K}}{T_\mathrm{K}}= \frac{3\alpha^{\ton{2}}}{2\mu a}, \, N=2 \label{eq:devano2}
\eeq

\beq
\frac{T_\mathrm{ano}-T_\mathrm{K}}{T_\mathrm{K}}= \frac{3\alpha^{\ton{3}}}{2\mu a^{2}}, \, N=3 \label{eq:devano3}
\eeq

\beq
\frac{T_\mathrm{dra}-T_\mathrm{K}}{T_\mathrm{K}}= \frac{2\alpha^{\ton{2}}}{\mu a}, \, N=2 \label{eq:devdra2}
\eeq

\beq
\frac{T_\mathrm{dra}-T_\mathrm{K}}{T_\mathrm{K}}= \frac{2\alpha^{\ton{3}}}{\mu a^{2}}, \, N=3 \label{eq:devdra3}
\eeq

\beq
\frac{T_\mathrm{sid}-T_\mathrm{K}}{T_\mathrm{K}}= \frac{2\alpha^{\ton{2}}}{\mu a}, \, N=2 \label{eq:devsid2}
\eeq

\beq
\frac{T_\mathrm{sid}-T_\mathrm{K}}{T_\mathrm{K}}= \frac{2\alpha^{\ton{3}}}{\mu a^{2}}, \, N=3 \label{eq:devsid3}
\eeq

From the above results, it is evident that the impact of the perturbation is more important for exoplanets orbiting in small orbits around a low-mass hosting star.
Moreover, in this ideal case, it is possible to build the quantities
\beq
\left(\frac{T_\mathrm{exp}-T_\mathrm{K}}{T_\mathrm{K}} \right)\mu a\,\,\,  \mathrm{for}\,\,\ N=2 \label{eq:un1N2}
\eeq

\beq
\left(\frac{T_\mathrm{exp}-T_\mathrm{K}}{T_\mathrm{K}} \right)\mu a^{2}\,\,\,  \mathrm{for}\,\,\ N=3 \label{eq:un1N3}
\eeq
which do not depend on the planetary system but  on the modified gravity model only. {For comparison, the lowest post-Newtonian corrections, as calculated in \cite{2016MNRAS.460.2445I}, lead to relative perturbations of the orbital period in the form
\beq
\frac{T_\mathrm{exp}-T_\mathrm{K}}{T_\mathrm{K}} \simeq \frac{R_{S}}{a} \label{eq:pk1pn}
\eeq
where $R_{S}$ is the Schwarzschild radius of the hosting star.}

 {Using the exoplanet catalogue\footnote{See, again \url{http://www.exoplanet.eu/catalog/} on the Internet.} we may evaluate the above quantities to obtain preliminary estimates for the model parameters $\alpha^{\ton{2}}, \alpha^{\ton{3}}$. Accordingly, if we consider the exoplanet Kepler-77b \citep{2013A&A...557A..74G}, $a=0.04501$ AU, $M=0.95\,  M_{\mathrm{Sun}}$ and supposing that $T_\mathrm{exp}-T_\mathrm{K}$ is of the order of the orbital period error amounting to $\sigma_{T_\mathrm{exp}}=2.3\times 10^{-7}\,\mathrm{d}=0.02\,\mathrm{s}$, we obtain as upper limit  $|\alpha^{\ton{2}}|\leq 10^{22}\, \mathrm{{m^{4}}\,{s^{-2}}}$. Similarly, if we consider the exoplanet WASP-12b \citep{2006PASP..118.1407P},  characterized by $a=0.0234$ AU, $M=1.434\,  M_{\mathrm{Sun}}$, $\sigma_{T_\mathrm{exp}}=1.44\times 10^{-7}\,\mathrm{d}=0.01\,\mathrm{s}$, we obtain  $|\alpha^{\ton{3}}|\leq 10^{32}\, \mathrm{{m^{5}}\,{s^{-2}}}$.}

{We may compare the above estimates to those obtained in \citet{Iorio:2018fam}, where   data records of the Earth's artificial satellites of the LAGEOS family were used. Taking into account the different parameterisation of the perturbing accelerations, the constraints for the $K_{2}, K_{3}$ parameters used in \citet{Iorio:2018fam}, since $\left|\alpha^{\ton 2}\right|=2 \left|K_{2}\right|$ and $\left|\alpha^{\ton 3}\right|= 6 \left|K_{3}\right|$,  imply   $\left|\alpha^{\ton{2}}\right|\leq 4.2\times 10^6\, \mathrm{{m^{4}}\,{s^{-2}}}$ and $\left|\alpha^{\ton{3}}\right|\leq 10^{13}\, \mathrm{{m^{5}}\,{s^{-2}}}$. Hence, we see that the preliminary constraints obtained from exoplanets are worse. {It is important to point out that, in general,  a comparison is actually meaningful only for those cases where the theory parameter $\alpha^{(N)}$ does not depend on the characteristics of the planetary system (mass, charge of host star) but on the gravity model only. This is the case, for instance of a $f(T)$ gravity model, which has extensively been studied by \citet{Iorio:2012cm,Xie:2013vua, Ruggiero:2015oka} by looking at the supplementary advances of the perihelia of some planets of our Solar system:  in this model a perturbation  in the form (\ref{eq:Wpos}) with $N=2$ arises.} In particular, if we consider our results, we may constrain the theory parameter $\alpha$ used in those works. Since it is $\left|\alpha^{\ton 2}\right|= 32 c^{2}\left|\alpha\right|$, we obtain $\left|\alpha\right| \leq 9.5\times 10^3\, \mathrm{m^{2}}$. It is interesting to notice that, if, on the one hand, this bound is worse than those obtained by \citet{Xie:2013vua} ($\left|\alpha\right| \leq 1.2 \times 10^2\, \mathrm{m^{2}}$)  and \citet{Ruggiero:2015oka} ($\left|\alpha\right| \leq 2.3 \times 10^1\, \mathrm{m^{2}}$) by about $\simeq 2-3$ orders of magnitude, on the other hand, it is better than, or at least comparable to, that obtained by \citet{Iorio:2012cm} ($\left|\alpha\right| \leq 1.8 \times 10^4\, \mathrm{m^{2}}$).}


\section{Conclusions} \label{sec:conc}
%
By analytically calculating the corrections to the anomalistic, draconitic and sideral periods due to the  perturbations of the Newtonian potential falling off as the square or the cube of the distance from the central mass, we have preliminarily laid the groundwork for testing such modified models of gravity also in planetary scenarios other than our solar system with the aim of the increasing amount of data available from transiting exoplanets.

While in the purely Keplerian case the aforementioned periods are degenerate, pK accelerations break the degeneracy and give rise to different expressions that we have analytically obtained. We remark that, such perturbations, are quite general and may derive from different modified gravity models, as discussed in \cite{Iorio:2018fam}.

{For those gravity models where the theory parameters are independent of the characteristics of the planetary system, such as the mass of the host star, our results show that the impact of these corrections is more important for planets orbiting in small orbits around low-mass stars.}

However, it is important to emphasize  that our results are just intended to yield preliminary insights on the potential of the method proposed in constraining the modified models of gravity considered if and when actual dynamical modeling will reach the experimental accuracy level, and to design sensitivity analyses. They should not be regarded as actual tests since, in this case, the data of the exoplanets of interest should be reprocessed with ad-hoc modified dynamical models including also the standard pK features of motion and the alternative model(s) one is interested in.

Indeed, the values of the currently available orbital and physical parameters which should be inserted in (\ref{ano2}) to (\ref{sid3}) come from purely Keplerian data analyses. As such, they are unavoidably a priori \virg{imprinted} by all the unmodelled pK dynamics, standard {(such as, for instance, terms in the form  (\ref{eq:pk1pn}))} and exotic, thus making the resulting values of the corrections $\Delta T$ of (\ref{ano2}) to (\ref{sid3}) purely indicative of the potentiality of the method.
As pointed out in \cite{2016MNRAS.460.2445I}, it would be important to find some exoplanetary systems for which it is possible to
measure independently measured two different orbital periods identifiable with some of the ones examined here. Indeed, their difference $\delta T $ would cancel out the common Keplerian component $T_\mathrm{K}$, which is a major source of systematic error because of the lingering uncertainty on the stellar mass, leaving just a pK correction. On the other hand, even in this case, some competing classical and general relativistic pK terms  may still enter $\delta T$, depending on the system's orbital geometry \cite{2016MNRAS.460.2445I}. A possible solution would be, at least in principle, a generalization of the method proposed in a different context \cite{1990grg..conf..313S} by setting up suitable linear combinations of as many different periods as possible for the same planet and/or of more planets belonging to the same extrasolar system in order to disentangle, by construction, $\alpha^{\ton{N}}$ from the other standard classical and relativistic pK corrections.

\bibliography{esop}

\begin{thebibliography}{26}
\expandafter\ifx\csname natexlab\endcsname\relax\def\natexlab#1{#1}\fi
\expandafter\ifx\csname bibnamefont\endcsname\relax
  \def\bibnamefont#1{#1}\fi
\expandafter\ifx\csname bibfnamefont\endcsname\relax
  \def\bibfnamefont#1{#1}\fi
\expandafter\ifx\csname citenamefont\endcsname\relax
  \def\citenamefont#1{#1}\fi
\expandafter\ifx\csname url\endcsname\relax
  \def\url#1{\texttt{#1}}\fi
\expandafter\ifx\csname urlprefix\endcsname\relax\def\urlprefix{URL }\fi
\providecommand{\bibinfo}[2]{#2}
\providecommand{\eprint}[2][]{\url{#2}}

\bibitem[{\citenamefont{Mayor and Queloz}(1995)}]{Mayor:1995eh}
\bibinfo{author}{\bibfnamefont{M.}~\bibnamefont{Mayor}} \bibnamefont{and}
  \bibinfo{author}{\bibfnamefont{D.}~\bibnamefont{Queloz}},
  \bibinfo{journal}{Nature} \textbf{\bibinfo{volume}{378}},
  \bibinfo{pages}{355} (\bibinfo{year}{1995}).

\bibitem[{\citenamefont{Perryman}(2018)}]{perryman}
\bibinfo{author}{\bibfnamefont{M.}~\bibnamefont{Perryman}},
  \emph{\bibinfo{title}{The exoplanet handbook}} (\bibinfo{publisher}{Cambridge
  University Press}, \bibinfo{year}{2018}).

\bibitem[{\citenamefont{{Deeg} and {Belmonte}}(20018)}]{book018}
\bibinfo{editor}{\bibfnamefont{H.~J.} \bibnamefont{{Deeg}}} \bibnamefont{and}
  \bibinfo{editor}{\bibfnamefont{J.~A.} \bibnamefont{{Belmonte}}}, eds.,
  \emph{\bibinfo{title}{Handbook of Exoplanets}} (\bibinfo{publisher}{Springer
  International Publishing}, \bibinfo{year}{20018}).

\bibitem[{\citenamefont{{Will}}(2015)}]{2014arXiv1409.7871W}
\bibinfo{author}{\bibfnamefont{C.~M.} \bibnamefont{{Will}}}, in
  \emph{\bibinfo{booktitle}{{General Relativity and Gravitation. A Centennial
  Perspective}}}, edited by
  \bibinfo{editor}{\bibfnamefont{A.}~\bibnamefont{{Ashtekar}}},
  \bibinfo{editor}{\bibfnamefont{B.~K.} \bibnamefont{{Berger}}},
  \bibinfo{editor}{\bibfnamefont{J.}~\bibnamefont{{Isenberg}}},
  \bibnamefont{and}
  \bibinfo{editor}{\bibfnamefont{M.}~\bibnamefont{{MacCallum}}}
  (\bibinfo{publisher}{Cambridge University Press, Cambridge},
  \bibinfo{year}{2015}), pp. \bibinfo{pages}{49--96}.

\bibitem[{\citenamefont{{Debono} and {Smoot}}(2016)}]{2016Univ....2...23D}
\bibinfo{author}{\bibfnamefont{I.}~\bibnamefont{{Debono}}} \bibnamefont{and}
  \bibinfo{author}{\bibfnamefont{G.~F.} \bibnamefont{{Smoot}}},
  \bibinfo{journal}{Universe} \textbf{\bibinfo{volume}{2}}, \bibinfo{eid}{23}
  (\bibinfo{year}{2016}), \eprint{1609.09781}.

\bibitem[{\citenamefont{Stelle}(1977)}]{Stelle:1976gc}
\bibinfo{author}{\bibfnamefont{K.~S.} \bibnamefont{Stelle}},
  \bibinfo{journal}{Phys. Rev.} \textbf{\bibinfo{volume}{D16}},
  \bibinfo{pages}{953} (\bibinfo{year}{1977}).

\bibitem[{\citenamefont{{Lake}}(2016)}]{2016Univ....2...24L}
\bibinfo{author}{\bibfnamefont{M.}~\bibnamefont{{Lake}}},
  \bibinfo{journal}{Universe} \textbf{\bibinfo{volume}{2}}, \bibinfo{pages}{24}
  (\bibinfo{year}{2016}).

\bibitem[{\citenamefont{Tsujikawa}(2010)}]{Tsujikawa:2010zza}
\bibinfo{author}{\bibfnamefont{S.}~\bibnamefont{Tsujikawa}},
  \bibinfo{journal}{Lect. Notes Phys.} \textbf{\bibinfo{volume}{800}},
  \bibinfo{pages}{99} (\bibinfo{year}{2010}), \eprint{1101.0191}.

\bibitem[{\citenamefont{{Clifton} et~al.}(2012)\citenamefont{{Clifton},
  {Ferreira}, {Padilla}, and {Skordis}}}]{2012PhR...513....1C}
\bibinfo{author}{\bibfnamefont{T.}~\bibnamefont{{Clifton}}},
  \bibinfo{author}{\bibfnamefont{P.~G.} \bibnamefont{{Ferreira}}},
  \bibinfo{author}{\bibfnamefont{A.}~\bibnamefont{{Padilla}}},
  \bibnamefont{and}
  \bibinfo{author}{\bibfnamefont{C.}~\bibnamefont{{Skordis}}},
  \bibinfo{journal}{Phys. Rep.} \textbf{\bibinfo{volume}{513}},
  \bibinfo{pages}{1} (\bibinfo{year}{2012}), \eprint{1106.2476}.

\bibitem[{\citenamefont{Berti et~al.}(2015)}]{Berti:2015itd}
\bibinfo{author}{\bibfnamefont{E.}~\bibnamefont{Berti}} \bibnamefont{et~al.},
  \bibinfo{journal}{Class. Quant. Grav.} \textbf{\bibinfo{volume}{32}},
  \bibinfo{pages}{243001} (\bibinfo{year}{2015}), \eprint{1501.07274}.

\bibitem[{\citenamefont{{Iorio} and {Ruggiero}}(2010)}]{2010OAJ.....3..167I}
\bibinfo{author}{\bibfnamefont{L.}~\bibnamefont{{Iorio}}} \bibnamefont{and}
  \bibinfo{author}{\bibfnamefont{M.~L.} \bibnamefont{{Ruggiero}}},
  \bibinfo{journal}{The Open Astronomy Journal} \textbf{\bibinfo{volume}{3}},
  \bibinfo{pages}{167} (\bibinfo{year}{2010}), \eprint{0909.5355}.

\bibitem[{\citenamefont{{Vargas dos Santos} and
  {Mota}}(2017)}]{2017PhLB..769..485V}
\bibinfo{author}{\bibfnamefont{M.}~\bibnamefont{{Vargas dos Santos}}}
  \bibnamefont{and} \bibinfo{author}{\bibfnamefont{D.}~\bibnamefont{{Mota}}},
  \bibinfo{journal}{Phys. Lett. B} \textbf{\bibinfo{volume}{769}},
  \bibinfo{pages}{485} (\bibinfo{year}{2017}), \eprint{1603.03243}.

\bibitem[{\citenamefont{Stairs}(2003)}]{Stairs2003}
\bibinfo{author}{\bibfnamefont{I.~H.} \bibnamefont{Stairs}},
  \bibinfo{journal}{Living Reviews in Relativity} \textbf{\bibinfo{volume}{6}},
  \bibinfo{pages}{5} (\bibinfo{year}{2003}), ISSN \bibinfo{issn}{1433-8351},
  \urlprefix\url{https://doi.org/10.12942/lrr-2003-5}.

\bibitem[{\citenamefont{Kramer et~al.}(2006)\citenamefont{Kramer, Stairs,
  Manchester, McLaughlin, Lyne, Ferdman, Burgay, Lorimer, Possenti, D'Amico
  et~al.}}]{kramer2006tests}
\bibinfo{author}{\bibfnamefont{M.}~\bibnamefont{Kramer}},
  \bibinfo{author}{\bibfnamefont{I.~H.} \bibnamefont{Stairs}},
  \bibinfo{author}{\bibfnamefont{R.}~\bibnamefont{Manchester}},
  \bibinfo{author}{\bibfnamefont{M.}~\bibnamefont{McLaughlin}},
  \bibinfo{author}{\bibfnamefont{A.}~\bibnamefont{Lyne}},
  \bibinfo{author}{\bibfnamefont{R.}~\bibnamefont{Ferdman}},
  \bibinfo{author}{\bibfnamefont{M.}~\bibnamefont{Burgay}},
  \bibinfo{author}{\bibfnamefont{D.}~\bibnamefont{Lorimer}},
  \bibinfo{author}{\bibfnamefont{A.}~\bibnamefont{Possenti}},
  \bibinfo{author}{\bibfnamefont{N.}~\bibnamefont{D'Amico}},
  \bibnamefont{et~al.}, \bibinfo{journal}{Science}
  \textbf{\bibinfo{volume}{314}}, \bibinfo{pages}{97} (\bibinfo{year}{2006}).

\bibitem[{\citenamefont{Blanchet et~al.}(2019)\citenamefont{Blanchet,
  H{\'e}brard, and Larrouturou}}]{Blanchet:2019zxv}
\bibinfo{author}{\bibfnamefont{L.}~\bibnamefont{Blanchet}},
  \bibinfo{author}{\bibfnamefont{G.}~\bibnamefont{H{\'e}brard}},
  \bibnamefont{and}
  \bibinfo{author}{\bibfnamefont{F.}~\bibnamefont{Larrouturou}},
  \bibinfo{journal}{Astron. Astrophys.} \textbf{\bibinfo{volume}{628}},
  \bibinfo{pages}{A80} (\bibinfo{year}{2019}), \eprint{1905.06630}.

\bibitem[{\citenamefont{{Winn}}(2010)}]{Winn010}
\bibinfo{author}{\bibfnamefont{J.~N.} \bibnamefont{{Winn}}}, in
  \emph{\bibinfo{booktitle}{{Exoplanets}}}, edited by
  \bibinfo{editor}{\bibfnamefont{S.}~\bibnamefont{{Seager}}}
  (\bibinfo{publisher}{University of Arizona Press, Tucson},
  \bibinfo{year}{2010}), pp. \bibinfo{pages}{55--77}.

\bibitem[{\citenamefont{{Iorio}}(2016)}]{2016MNRAS.460.2445I}
\bibinfo{author}{\bibfnamefont{L.}~\bibnamefont{{Iorio}}},
  \bibinfo{journal}{MNRAS} \textbf{\bibinfo{volume}{460}},
  \bibinfo{pages}{2445} (\bibinfo{year}{2016}), \eprint{1407.5021}.

\bibitem[{\citenamefont{Ruggiero}(2014)}]{SSS}
\bibinfo{author}{\bibfnamefont{M.~L.} \bibnamefont{Ruggiero}},
  \bibinfo{journal}{Int.J.Mod.Phys.} \textbf{\bibinfo{volume}{D23}},
  \bibinfo{pages}{1450049} (\bibinfo{year}{2014}), \eprint{1010.2114}.

\bibitem[{\citenamefont{{Brumberg}}(1991)}]{1991ercm.book.....B}
\bibinfo{author}{\bibfnamefont{V.~A.} \bibnamefont{{Brumberg}}},
  \emph{\bibinfo{title}{{Essential Relativistic Celestial Mechanics}}}
  (\bibinfo{publisher}{Adam Hilger, Bristol}, \bibinfo{year}{1991}).

\bibitem[{\citenamefont{{Gandolfi} et~al.}(2013)\citenamefont{{Gandolfi},
  {Parviainen}, {Fridlund}, {Hatzes}, {Deeg}, {Frasca}, {Lanza}, {Prada
  Moroni}, {Tognelli}, {McQuillan} et~al.}}]{2013A&A...557A..74G}
\bibinfo{author}{\bibfnamefont{D.}~\bibnamefont{{Gandolfi}}},
  \bibinfo{author}{\bibfnamefont{H.}~\bibnamefont{{Parviainen}}},
  \bibinfo{author}{\bibfnamefont{M.}~\bibnamefont{{Fridlund}}},
  \bibinfo{author}{\bibfnamefont{A.~P.} \bibnamefont{{Hatzes}}},
  \bibinfo{author}{\bibfnamefont{H.~J.} \bibnamefont{{Deeg}}},
  \bibinfo{author}{\bibfnamefont{A.}~\bibnamefont{{Frasca}}},
  \bibinfo{author}{\bibfnamefont{A.~F.} \bibnamefont{{Lanza}}},
  \bibinfo{author}{\bibfnamefont{P.~G.} \bibnamefont{{Prada Moroni}}},
  \bibinfo{author}{\bibfnamefont{E.}~\bibnamefont{{Tognelli}}},
  \bibinfo{author}{\bibfnamefont{A.}~\bibnamefont{{McQuillan}}},
  \bibnamefont{et~al.}, \bibinfo{journal}{Astron. Astrophys.}
  \textbf{\bibinfo{volume}{557}}, \bibinfo{eid}{A74} (\bibinfo{year}{2013}),
  \eprint{1305.3891}.

\bibitem[{\citenamefont{{Pollacco} et~al.}(2006)\citenamefont{{Pollacco},
  {Skillen}, {Collier Cameron}, {Christian}, {Hellier}, {Irwin}, {Lister},
  {Street}, {West}, {Anderson} et~al.}}]{2006PASP..118.1407P}
\bibinfo{author}{\bibfnamefont{D.~L.} \bibnamefont{{Pollacco}}},
  \bibinfo{author}{\bibfnamefont{I.}~\bibnamefont{{Skillen}}},
  \bibinfo{author}{\bibfnamefont{A.}~\bibnamefont{{Collier Cameron}}},
  \bibinfo{author}{\bibfnamefont{D.~J.} \bibnamefont{{Christian}}},
  \bibinfo{author}{\bibfnamefont{C.}~\bibnamefont{{Hellier}}},
  \bibinfo{author}{\bibfnamefont{J.}~\bibnamefont{{Irwin}}},
  \bibinfo{author}{\bibfnamefont{T.~A.} \bibnamefont{{Lister}}},
  \bibinfo{author}{\bibfnamefont{R.~A.} \bibnamefont{{Street}}},
  \bibinfo{author}{\bibfnamefont{R.~G.} \bibnamefont{{West}}},
  \bibinfo{author}{\bibfnamefont{D.~R.} \bibnamefont{{Anderson}}},
  \bibnamefont{et~al.}, \bibinfo{journal}{PASP} \textbf{\bibinfo{volume}{118}},
  \bibinfo{pages}{1407} (\bibinfo{year}{2006}), \eprint{astro-ph/0608454}.

\bibitem[{\citenamefont{Iorio and Ruggiero}(2018)}]{Iorio:2018fam}
\bibinfo{author}{\bibfnamefont{L.}~\bibnamefont{Iorio}} \bibnamefont{and}
  \bibinfo{author}{\bibfnamefont{M.~L.} \bibnamefont{Ruggiero}},
  \bibinfo{journal}{JCAP} \textbf{\bibinfo{volume}{1810}}, \bibinfo{pages}{021}
  (\bibinfo{year}{2018}), \eprint{1807.11807}.

\bibitem[{\citenamefont{Iorio and Saridakis}(2012)}]{Iorio:2012cm}
\bibinfo{author}{\bibfnamefont{L.}~\bibnamefont{Iorio}} \bibnamefont{and}
  \bibinfo{author}{\bibfnamefont{E.~N.} \bibnamefont{Saridakis}},
  \bibinfo{journal}{Mon. Not. Roy. Astron. Soc.}
  \textbf{\bibinfo{volume}{427}}, \bibinfo{pages}{1555} (\bibinfo{year}{2012}),
  \eprint{1203.5781}.

\bibitem[{\citenamefont{Xie and Deng}(2013)}]{Xie:2013vua}
\bibinfo{author}{\bibfnamefont{Y.}~\bibnamefont{Xie}} \bibnamefont{and}
  \bibinfo{author}{\bibfnamefont{X.-M.} \bibnamefont{Deng}},
  \bibinfo{journal}{Mon. Not. Roy. Astron. Soc.}
  \textbf{\bibinfo{volume}{433}}, \bibinfo{pages}{3584} (\bibinfo{year}{2013}),
  \eprint{1312.4103}.

\bibitem[{\citenamefont{Ruggiero and Radicella}(2015)}]{Ruggiero:2015oka}
\bibinfo{author}{\bibfnamefont{M.~L.} \bibnamefont{Ruggiero}} \bibnamefont{and}
  \bibinfo{author}{\bibfnamefont{N.}~\bibnamefont{Radicella}},
  \bibinfo{journal}{Phys. Rev.} \textbf{\bibinfo{volume}{D91}},
  \bibinfo{pages}{104014} (\bibinfo{year}{2015}), \eprint{1501.02198}.

\bibitem[{\citenamefont{{Shapiro}}(1990)}]{1990grg..conf..313S}
\bibinfo{author}{\bibfnamefont{I.~I.} \bibnamefont{{Shapiro}}}, in
  \emph{\bibinfo{booktitle}{General Relativity and Gravitation, 1989}}, edited
  by \bibinfo{editor}{\bibfnamefont{N.}~\bibnamefont{{Ashby}}},
  \bibinfo{editor}{\bibfnamefont{D.~F.} \bibnamefont{{Bartlett}}},
  \bibnamefont{and} \bibinfo{editor}{\bibfnamefont{W.}~\bibnamefont{{Wyss}}}
  (\bibinfo{publisher}{Cambridge University Press, Cambridge},
  \bibinfo{year}{1990}), pp. \bibinfo{pages}{313--330}.

\end{thebibliography}


\begin{thebibliography}{200}

\end{thebibliography}

\end{document}